SHORT PAPER

# Computer Self-efficacy and Its Relationship with Web Portal Usage: Evidence from the University of the East

Rex P. Bringula
University of the East
(corresponding author)

Julius Jan M. Sarmiento
University of the East

Roselle S. Basa
University of the East

**Abstract**

*Purpose* – The University of the East Web Portal is an academic, web-based system that provides educational electronic materials and e-learning services. To fully optimize its usage, it is imperative to determine the factors that relate to its usage. Thus, this study, to determine the computer self-efficacy of the faculty members of the University of the East and its relationship with their web portal usage, was conceived.

*Method* – Using a validated questionnaire, the profile of the respondents, their computer self-efficacy, and web portal usage were gathered.

*Results* – Data showed that the respondents were relatively young ($M$ = 40 years old), majority had master's degree (f = 85, 72%), most had been using the web portal for four semesters (f = 60, 51%), and the large part were intermediate web portal users (f = 69, 59%). They were highly skilled in using the computer ($M$ = 4.29) and skilled in using the Internet ($M$ = 4.28). E-learning services ($M$ = 3.29) and online library resources ($M$ = 3.12) were only used occasionally. Pearson correlation revealed that age was positively correlated with online library resources ($r$ = 0.267, $p < 0.05$) and a negative relationship existed between perceived skill level in using the portal and online library resources usage ($r$ = -0.206, $p < 0.05$). A 2x2 $\chi^2$ revealed that the highest educational attainment had a significant relationship with online library resources ($\chi^2$ = 5.489, df = 1, $p < 0.05$). Basic computer ($r$ = 0.196, $p < 0.05$) and Internet skills ($r$ = 0.303, $p < 0.05$) were significantly and positively related with e-learning services usage but not with online library resources usage.

*Research Implication* – Other individual factors such as attitudes towards the web portal and anxiety towards using the web portal can be investigated.

# INTRODUCTION

Educational web portals serve as gateways to information and services for learning or teaching (Manouselis et al., 2009). Universities around the world recognize the power of web portals when they develop their own web portals. The University of the East has four web portals, namely, administrative, alumni, student, and faculty. Each portal has its own functions and capabilities. The faculty portal is supporting e-learning services (e.g., posting of announcements, uploading of files, downloading of classlists, etc.) and providing online library resources (e.g., e-books, e-journals, cases, magazines, etc.). The perceived importance that the portal can provide prompted the researchers to study its usability (e.g., Bringula & Basa, 2011).

However, despite the rich literature on computer self-efficacy, no studies were conducted to study its role with the web faculty portal usage. This study wants to address this gap. It sought answers to the following questions. 1) What is the profile of the respondents in terms of age, highest educational attainment, length of use of the web portal, and perceived skill level in using the portal? 2) What is the level of computer self-efficacy of the respondents in terms of basic computer and Internet skills? 3) What is the web portal usage of the respondents with regard to e-learning services and online library resources? 4) Is there a significant relationship between the profile of the respondents and their web portal usage? and 5) Is there a significant relationship between computer self-efficacy and web portal usage?

It is hypothesized that a) there is no significant relationship between the profile of the respondents and their web portal usage, and b) there is no significant relationship between computer self-efficacy and web portal usage.

# LITERATURE REVIEW

*Self-efficacy*

According to the self-efficacy theory, a person could perform a task because that person believes he or she is capable of achieving a certain outcome (Maddux, 2007). A person's capacity to respond appropriately to a situation depends on his or her self-efficacy (Maddux, 2007). Thus, self-efficacy beliefs are task specific (Palladino, 2006), subjective, and situation-dependent (Maddux, 2007). Studies were conducted to determine the significance of self-efficacy in various aspects.

Fagan, Neill, and Wooldridge (2003) investigated the factors perceived to affect an individual's use of Information Technology. The study was based on Bandura's Social Cognitive Theory (SCT) and Triandis' Theory of Interpersonal Behavior (TIB). As suggested by SCT and from the data collected from 978 respondents, experience and support were positively related to computer self-efficacy, and computer self-efficacy was negatively related to anxiety and positively related to usage.

Zhang, Prybutok, and Huang (2006) disclosed that the skills and experiences in using a computer of users have positive correlation with e-service satisfaction. Prior internet experience influences the ability of a person to perform specific tasks using the Internet (Alenezi, Abdul Karim, & Veloo, 2010). Internet users may be classified as novice, expert, or professional (Banati, Bedi, & Grover, 2006).

Chien (2012) investigated the influence of system and instructor factors on e-learning effectiveness under the interactions of computer self-efficacy. E-learning effectiveness was measured in terms of functionality, interaction, and response while e-learning instructor factors were attitude, technical skills, and instructional method. The moderating effects of learners' computer self-efficacy were examined. Responses from the three hundred fourteen (314) respondents from the financial services industry in Taiwan were statistically analyzed and showed that both system and instructor factors had significant positive influence on e-learning effectiveness, and that learners' computer self-efficacy had a moderating effect on the relationship between system functionality and training effectiveness. The higher the computer self-efficacy, the stronger the relationship was between functionality and e-learning effectiveness, and vice versa.

*Web Portal*

Xia (2003) found that students and teachers prefer broad collection of updated library resources that are convenient and easy to access. It was shown that students are fond of digital library collections because they save time (Liu & Luo, 2011). Hence, they want to access these educational resources anytime of the day at their convenient place (Liu & Luo, 2011). The study of Heinrichs, Lim, Lim, and Spangenberg (2007) found that satisfaction and higher usage of academic library were influenced by the design of the website. In order to materialize the increased utilization of library collections, the authors suggested that library administrators ensure that the design factors be incorporated in the website. The study concluded that proper academic library management is needed for the improvement of library services.

The study of Santosa (2009) also conducted a similar study on Papirus, an e-learning system developed in one department of a big university wherein students' attitude towards e-learning and how certain usability factors affected students' attitude towards e-learning were determined. It was found out that ease of navigation had a strong effect on perceived ease of use, which in turn affected user attitude and satisfaction. The study of Bringula and Basa (2011) investigated the factors that might affect web portal usability. Using multiple regression analysis on the data gathered from 82 faculty members, it was revealed that information content (e.g., teaching materials provided by the portal and the mechanisms of disseminating these materials) was the only significant predictor of web portal usability.

## METHODOLOGY

This descriptive study utilized a pilot-tested and validated questionnaire. Basic computer skills (all factor loadings (f.l.) > 0.70; Cronbach's alpha = 0.981), Internet computer skills (all f.l. > 0.70; Cronbach's alpha = 0.962), E-learning services (f.l. > 0.70; Cronbach's alpha = 0.756), and Online Library Resources (f.l. > 0.70; Cronbach's alpha = 0.952) (see Bringula & Basa, 2011) were all found valid and reliable. The study was conducted at the University of the East during the second semester of school year 2010-2011. During this period, there were 725 full- and part-time faculty members of the six colleges of the University. Using Sloven's formula (e = 0.10), a minimum sample size of 88 was derived. To ensure a suitable number of responses, three hundred fifty-three (353) forms were distributed. One hundred eighteen (118) forms were retrieved and were all used in the analysis.

The questionnaire was composed of three parts. The first part gathered faculty profile in terms of age, highest educational attainment (undergraduate or further studies), length of use of the web portal (0 – never used the portal to 7 – used the portal for at least more than 4 semesters), and perceived skill level in using the portal (e.g., beginner, intermediate, or expert). The second part allowed the respondents to rate their basic computer and Internet skills from Not Skilled (1) to Highly Skilled (5). Lastly, the third part gathered their web portal usage (1 – Never to 5 – Very Often).

Frequency counts, mean, and percentage were utilized to describe the data. Pearson correlation and 2x2 $\chi^2$ were employed to test the hypotheses. A 5% level of probability and 95% reliability were employed to determine the significance of the findings.

## RESULTS AND DISCUSSION

Findings revealed that the average age of the respondents was 40. This showed that most of the respondents were relatively young. It was also found that majority had master's degree (f = 85, 72%). Respondents were aware of the web portal since all of them had been using the web portal. Notably, majority had been using the web portal for four semesters (f = 60, 51%) and they perceived that their skill level in using the portal as intermediate (f = 69, 59%). They were highly skilled in using the computer ($M$ = 4.29) and the Internet ($M$ = 4.28). These findings disclosed that the respondents had a certain level of familiarity with the system. Furthermore, their basic computer and Internet skills showed that they had the capabilities of using the web portal.

In terms of faculty web portal usage, e-learning services ($M$ = 3.29) and online library resources ($M$ = 3.12) were only used occasionally. The low usage of the faculty web portal is quite surprising considering the functionality, availability, and accessibility of its services and digital library resources. This is similar to the findings of Bringula and Basa (2011) which showed that the low usage of the faculty web portal could be attributed to two factors. First, the functionalities of the e-learning services (e.g., downloading of classlists, posting of announcements, etc.), though important, do not reinforce frequent use of the web portal. Second, they got used to the same materials that they have been using every semester. Hence, their attitude towards the use of the web portal may not be favorable. Further studies can shed light on this unexplored gap.

Meanwhile, Table 1 shows the correlation results of the variables. As shown in Table 1, age was positively related with online library resources ($r = 0.267$, $p < 0.05$) while perceived skill level in using the portal had a negative correlation with online library resources ($r = -0.206$, $p < 0.05$). It is worthy to note that age has positive relationship with online library resources. This reveals that as the respondents of this study get older, their online library usage increases. It can be argued that older people tend to appreciate the convenience brought by the technology in the context of online library resources. It is easier to search for teaching materials in the convenience of their own homes. In the past, teachers had to go to the library and search for voluminous records of materials just to find a suitable content for their class. Nowadays, teachers search these materials in just one click of the mouse. However, as the respondents become more skilled in using computers, they tend to get frustrated to the design of the online library resources. In return, the usability of the online web resources and its usage are compromised.

A 2x2 $\chi^2$ was carried out to determine whether there was a significant relationship between highest educational attainment and web portal elements. In terms of relationship between highest educational attainment and e-learning services, the relationship was not significant ($\chi^2 = 0.711$, df = 1, $p > 0.05$). On the other hand, there was a significant relationship between highest educational attainment and online library resources usage ($\chi^2 = 5.489$, df = 1, $p < 0.05$). This means that as the level of education of the respondents increase, it can be expected that their online library resources usage would also increase. This can be explained by the fact that further studies require numerous scholarly materials which is a good indication that online library resources of the web portal satisfy such needs.

Table 1. Relationship of Variables

| Variables | | E-learning Services | Online Library Resources |
|---|---|---|---|
| **Faculty Profile** | | | |
| Age | Pearson correlation ($r$) | 0.037 | 0.267 |
| | $p$-value | 0.692 | 0.004 |
| Highest Educational Attainment | $\chi^2$ (df = 1) | 0.711 | 5.489 |
| | $p$-value | 0.399 | 0.019 |
| Length of Use of the Web Portal | Pearson correlation ($r$) | 0.119 | -0.051 |
| | $p$-value | 0.200 | 0.583 |
| Perceived Skill Level in Using the Portal | Pearson correlation ($r$) | 0.124 | -0.206 |
| | $p$-value | 0.180 | 0.025 |
| **Computer Self-Efficacy** | | | |
| Basic Computer Skills | Pearson correlation ($r$) | 0.196 | 0.057 |
| | $p$-value | 0.034 | 0.539 |
| Internet Skills | Pearson correlation ($r$) | 0.303 | 0.176 |
| | $p$-value | 0.001 | 0.057 |

Basic computer skills ($r = 0.196$, $p < 0.05$) and Internet skills ($r = 0.303$, $p < 0.05$) were found to have significant and positive relationship with e-learning services usage but not with online library resources usage. Basic computer skills are skills that pertain to operating a computer, using a word processor, and manipulating files while Internet skills refer to the skills in using the World Wide Web. Thus, these skills are mandatory in order to use the e-learning services. This confirms the studies of Zhang et al. (2006) and Chien (2012).

There is an interesting pattern of correlation as can be gleaned from Table 2. Two factors of the faculty profile had relationship with online library resources. On the other hand, all variables on computer self-efficacy had correlation on e-learning services. This correlation pattern can be explained by the nature of the two web portal elements. It must be noted that online library resources are collection of electronic copies of journals, books, magazines, government reports, and other reference materials which can provide the pedagogical needs of the teachers, the direct beneficiaries of these materials. Thus, as shown in the correlation results, the use of these materials depends on their profile.

On the other hand, both variables of computer self-efficacy had positive and significant relationship with e-learning services. Similarly, this pattern can be explained by the nature of the e-learning services. The e-learning services entail posting of announcements, uploading of course materials, forum discussions, downloading of classlists, etc. Clearly, these online activities require computer and Internet skills in order to carry out these tasks. While searching online library resources can be done in the library with the assistance of a librarian, e-learning services should be done personally by the faculty due to confidential and security reasons. Thus, faculty members should have basic computer and Internet skills in order to facilitate e-learning services. Interestingly, no faculty profile variables were related to e-learning services. This implies that the use of e-learning services could be utilized regardless of age, highest educational attainment, familiarity with the web portal, and skill level.

## CONCLUSIONS AND RECOMMENDATIONS

The null hypotheses stating that a) there is no significant relationship between the profile of the respondents and their web portal usage, and b) there is no significant relationship between computer self-efficacy and web portal usage are both partially rejected. Computer self-efficacy influences e-learning services while three factors of faculty profile influence online library resources. These findings are attributed to the nature of the web portal services. Thus, it is concluded that the use of one element of the web portal can be influenced by one set of distinct factors.

It is recommended, therefore, that programs be conducted to optimize the use of this technology such as those that enhance the technical skills and level of awareness on the use of the web portal of the respondents. This study can be extended in two ways. First, the role of teachers' attitudes towards the web portal could be investigated. Second, the aspects of importance of the web portal and satisfaction on the use of the web portal could also be explored.

## ACKNOWLEDGEMENT

The researchers would like to thank the participants of the study. This is a revised paper which was presented in the International Conference on eLearning 2015.